\documentclass[aps,pra,twocolumn,superscriptaddress]{revtex4-1}
\usepackage{graphicx}
\usepackage{amssymb}
\usepackage{amsmath}
\usepackage{bm}
\usepackage{xcolor}

\begin{document}

\title{Critical Energy Dissipation in a Binary Superfluid Gas by a Moving Magnetic Obstacle}

\author{Joon Hyun Kim}
\affiliation{Department of Physics and Astronomy, and Institute of Applied Physics, Seoul National University, Seoul 08826, Korea}

\author{Deokhwa Hong}
\affiliation{Department of Physics and Astronomy, and Institute of Applied Physics, Seoul National University, Seoul 08826, Korea}
\affiliation{Center for Correlated Electron Systems, Institute for Basic Science, Seoul 08826, Korea}

\author{Kyuhwan Lee}
\affiliation{Department of Physics and Astronomy, and Institute of Applied Physics, Seoul National University, Seoul 08826, Korea}
\affiliation{Center for Correlated Electron Systems, Institute for Basic Science, Seoul 08826, Korea}

\author{Y. Shin}
\email{yishin@snu.ac.kr}
\affiliation{Department of Physics and Astronomy, and Institute of Applied Physics, Seoul National University, Seoul 08826, Korea}
\affiliation{Center for Correlated Electron Systems, Institute for Basic Science, Seoul 08826, Korea}


\begin{abstract}

We study the critical energy dissipation in an atomic superfluid gas with two symmetric spin components by an oscillating magnetic obstacle. Above a certain critical oscillation frequency, spin-wave excitations are generated by the magnetic obstacle, demonstrating the spin superfluid behavior of the system. When the obstacle is strong enough to cause density perturbations via local saturation of spin polarization, half-quantum vortices (HQVs) are created for higher oscillation frequencies, which reveals the characteristic evolution of critical dissipative dynamics from spin-wave emission to HQV shedding. Critical HQV shedding is further investigated using a pulsed linear motion of the obstacle, and we identify two critical velocities to create HQVs with different core magnetization.

\end{abstract}

\maketitle

Spin superfluidity, the absence of energy dissipation in a spin current, is a fascinating macroscopic quantum phenomenon. It was first observed in liquid $^3$He~\cite{Mukharskii87} and recently investigated in various magnetic materials~\cite{Hillebrands16,Han18,Lau18}, suggesting its potential applications in spintronics~\cite{Baltz18}. One minimal setting allowing the remarkable phenomenon is a binary superfluid system, which consists of two symmetric superflowing components. Owing to the $\mathbb{Z}_2$ symmetry, the system has two Goldstone modes corresponding to pure phonons and magnons~\cite{Ueda12}, which are associated with mass and spin superfluidity, respectively~\cite{Sonin10,Duine17}. In cold atom experiments, such a symmetric binary superfluid system was realized with spin-1 antiferromagnteic Bose-Einstein condensates (BECs) of $^{23}$Na. Its spin superfluid behavior was demonstrated by observing the absence of damping in spin dipole oscillations of trapped samples~\cite{Kim17,Ferrari18}. Two sound modes in the mass and spin sectors were also observed~\cite{Kim20}. 

One of the key characteristics of a superfluid is the critical velocity for its frictionless flow against external perturbations. In a conventional scalar superfluid with broken $U(1)$ symmetry, it is known that when it flows past an obstacle, energy dissipation occurs above a certain critical velocity via phonon radiation~\cite{Pitaevskii04} and nucleation of vortices~\cite{Varoquaux15}, arising from the local accumulation of superfluid phase slippages~\cite{Adams98}. An interesting question about a spin superfluid is how it responds to a moving {\it magnetic} obstacle, i.e., an obstacle that induces different perturbations to each spin component~\cite{Jung21}. Based on the analogy between the mass and spin sectors, it is expected that magnon excitations would be generated above a certain critical velocity. However, the situation is different for vortex nucleation because its fundamental topological excitations are vortices with fractional circulation, which are called half-quantum vortices (HQVs)~\cite{Seo15}. An HQV contains both mass and spin circulations, and therefore, its nucleation cannot be fulfilled by a pure phase slip process in the spin sector. 

In this Letter, we investigate the critical dissipative dynamics in a symmetric binary superfluid by an oscillating magnetic obstacle. Pertaining to a weak obstacle, which does not saturate local spin polarization, a sudden onset of spin-wave excitations is observed with increasing the oscillation frequency, which demonstrates the spin superfluidity of the system. Surprisingly, the creation of HQVs is not observed for the weak magnetic obstacle whose speed exceeds the spin sound velocity. On the other hand, with a strong magnetic obstacle, which can produce mass density perturbations by inducing local saturation of spin polarization, HQVs can be created by moving the obstacle above a certain critical velocity. Furthermore, we find that the critical velocities are different for the two types of HQVs with different core magnetizations, which originate from the magnetic property of the obstacle. This study demonstrates spin superfluid behavior of a binary superfluid system against external magnetic perturbations, and furthermore, reveals the evolution of critical dissipative dynamics from spin-wave emission to HQV shedding in a spin superfluid.

Our experiment starts with a BEC of $^{23}$Na in the $|F$$=$$1,m_{F}$$=$$0\rangle$ hyperfine ground state in an optical dipole trap~\cite{Kim17}. The condensate contains about $2.7 \times10^{6}$ atoms and its Thomas-Fermi radii are $(R_{x},R_{y},R_{z})$~$\approx$~$(162,106,1.5)~\mu$m for trapping frequencies of $(\omega_{x},\omega_{y},\omega_{z})=2\pi\times(5.8,8.9,641)$~Hz. We prepare an equal mixture of atoms in the two spin states, $|$$\uparrow\rangle \equiv |m_F$$=$$1\rangle$ and $|$$\downarrow\rangle \equiv |m_F$$=$$-1\rangle$, by applying a $\pi/2$ radio-frequency (rf) pulse to the initial $|m_F$$=$$0\rangle$ state. The two spin components are miscible~\cite{Ketterle98} and constitute a symmetric binary superfluid. The intercomponent interaction strength, $g_{\uparrow\downarrow}$, is comparable to the intracomponent interaction strength, $g$, given as $(g-g_{\uparrow\downarrow})/g \approx 7\%$~\cite{Tiemann11}, so the mass and spin sectors of the binary system are energetically well separated. For the peak atomic density at the condensate center, the density and spin healing lengths are $\xi_n\approx0.5~\mu$m and $\xi_s\approx2.5~\mu$m, respectively, and the speed of spin sound is $c_s=0.63(4)~$mm/s in the highly oblate condensate~\cite{Kim20}. During the experiment, spin-changing collisions are suppressed by a large negative quadratic Zeeman energy via microwave field dressing~\cite{Bloch06}. The external magnetic field is $50$~mG, and its gradient on the $xy$ plane is canceled to be less than $0.1$~mG/cm~\cite{Kim19}.

\begin{figure} [t]
	\includegraphics[width=8.5cm]{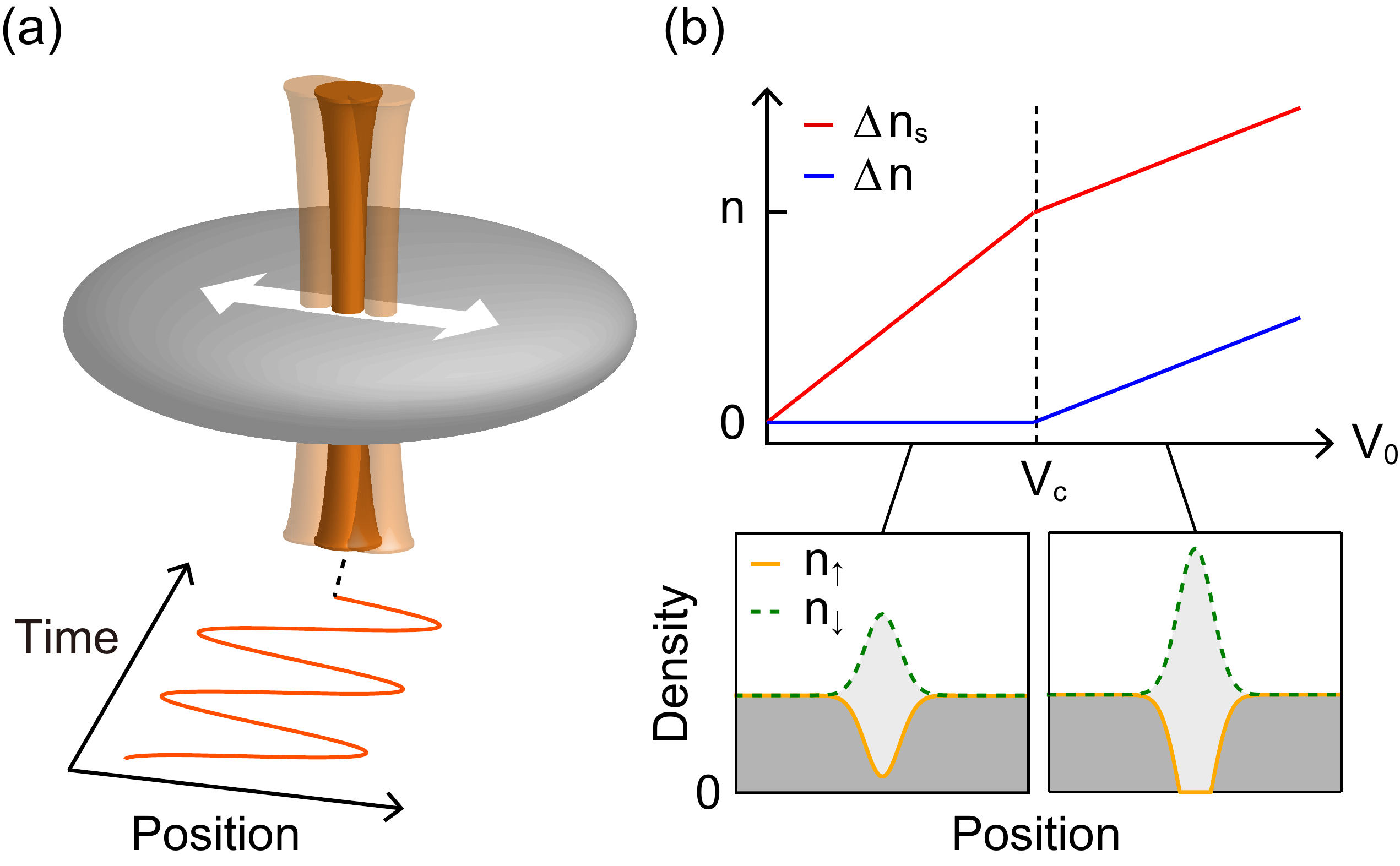}
	\caption{Oscillating magnetic obstacle in a symmetric binary superfluid. (a) Schematic of the experiment. A focused near-resonant Gaussian laser beam, which provides a repulsive (attractive) potential for the spin--$\uparrow$($\downarrow$) component, undergoes sinusoidal oscillations along a linear path at the central region of the trapped sample. (b) Spin and mass density variations, $\Delta n_s$ and $\Delta n$, as functions of the obstacle strength $V_0$. For $V_0>V_c$, the spin polarization is saturated due to the density depletion of the spin--$\uparrow$ component. The insets display the representative density profiles of the spin--$\uparrow$ (yellow solid) and spin--$\downarrow$ (green dashed) components for weak (left) and strong obstacles (right).
	}
\end{figure}

The schematic of our experiment is shown in Fig.~1(a). A magnetic obstacle is realized using a focused 589-nm near-resonant laser beam with circular polarization, which produces repulsive and attractive Gaussian optical potentials for the two $|$$\uparrow\rangle$ and $|$$\downarrow\rangle$ states, respectively, with same peak magnitude $V_0$~\cite{Kim20}. The beam propagates toward the central region of the condensate along the $z$ axis and its $1/e^2$ radius is about $7 \xi_s$. We adiabatically ramp up the obstacle beam for $300~$ms and hold it for $100~$ms to stabilize the beam intensity. Then, we sinusoidally oscillate the obstacle by manipulating a piezodriven mirror for 1~s at variable oscillation frequency $f$. The obstacle position is given by $x(t)=A\cos (2\pi f t)$ with $x=0$ denoting the sample center. The sweep distance is $2A \approx37~\mu$m, over which the atomic column density varies less than $ 5\%$. After the stirring process, we ramp down the obstacle beam for $300~$ms and take a spin-sensitive phase-contrast image of the sample along the $z$ direction to measure the spatial magnetization distribution~\cite{Seo15}. We let the condensate expand for 19~ms before applying the imaging light, which facilitates the observation of magnon excitations via their self-interference effect~\cite{Seo14} as well as HQVs with their expanded ferromagnetic cores~\cite{Seo15}.

Perturbations generated by the magnetic obstacle depend on the obstacle strength $V_0$. Figure 1(b) shows the spin and mass density variations, $\Delta n_s$ and $\Delta n$, induced at the center of a stationary obstacle as a function of $V_0$, where $\Delta n_s\equiv n_{\downarrow}-n_{\uparrow}$ and  $\Delta n\equiv n_{\downarrow}+n_{\uparrow}-n$ with $n_{\uparrow(\downarrow)}$ being the density of the spin--$\uparrow(\downarrow)$ component and $n$ being the total density without the obstacle. When $V_0$ is small, the density profiles of the two spin components vary antisymmetrically, yielding $\Delta n_s=2V_0/(g-g_{\uparrow\downarrow})$ with $\Delta n=0$, i.e., only spin perturbations are generated by the magnetic obstacle. However, when $V_0$ is increased over a certain critical strength $V_c$, the spin--$\uparrow$ component is locally depleted, resulting in $\Delta n>0$, and thus, mass perturbations are also induced by the magnetic obstacle. The critical strength is given by $V_c = (g-g_{\uparrow\downarrow})n/2$ from  $\Delta n_s = n$ and in our experiment, $V_c/\mu \approx 3.5\%$ with $\mu=(g+g_{\uparrow\downarrow})n/2$ being the chemical potential of the condensate. In the following, we call a magnetic obstacle with $V_0/V_c$$<$1 ($>$1) weak (strong).

\begin{figure} [t]
	\includegraphics[width=8.5cm]{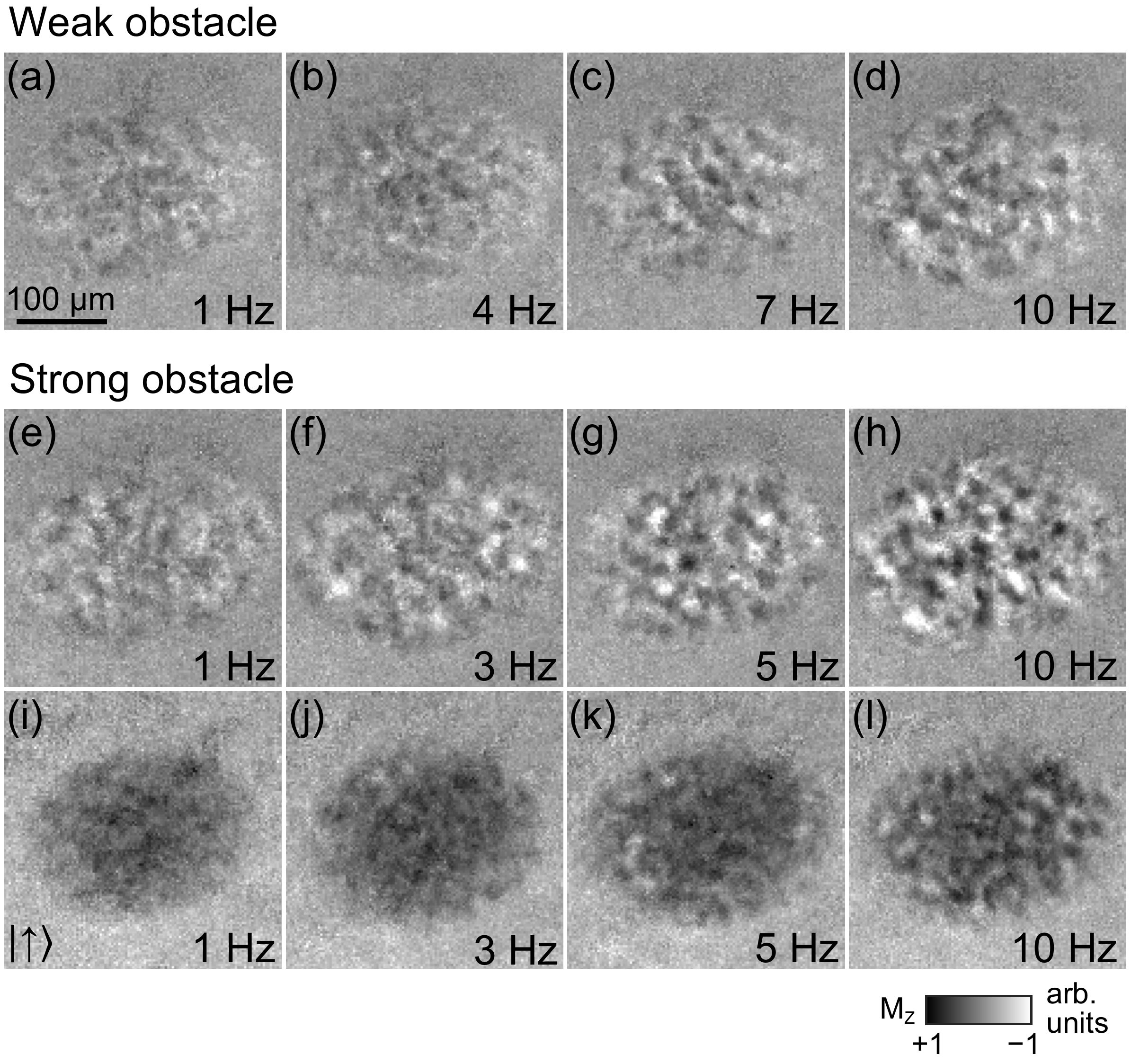}
	\caption{Generation of spin excitations in a spinor Bose-Einstein condensate (BEC) by an oscillating magnetic obstacle.  Magnetization ($M_z$) images of the condensate stirred with an obstacle of $V_0/V_c \approx 0.9$ [(a)-(d)] and $\approx 2.2$ [(e)-(h)] for various oscillation frequencies $f$. The images were obtained after a 19-ms time-of-flight. Fully magnetized pointlike domains in (g) and (h) indicate half-quantum vortices (HQVs) with ferromagnetic cores. (i)-(l) Images of the spin--$\uparrow$ component for the same stirring conditions in (e)-(h), taken after Stern-Gerlach spin separation. The HQVs are distinguishable as density-depleted holes in (k) and (l).
	}
\end{figure}

In Fig.~2, we display a series of magnetization images of the perturbed condensate for various stirring frequencies $f$ with weak and strong magnetic obstacles of $V_0/V_c\approx 0.9$ and 2.2, respectively. As $f$ increases, spin fluctuations in the condensate are observed to be enhanced, indicating that energy dissipation occurs by the oscillating magnetic obstacle. In the case of a strong obstacle, it is noticeable that fully spin-polarized, pointlike domains appear in the condensate at high $f >3$~Hz~[Figs.~2(g) and 2(h)]. This implies that HQVs are generated by the fast moving obstacle, and it is confirmed by taking an image of the sample after Stern-Gerlach  spin separation and observing the appearance of density-depleted holes in each spin component~[Figs.~2(k) and 2(l)]. By contrast, we observe that HQVs are not created with the weak obstacle for the full range of $f$ in our experiment, where the obstacle's speed reaches over $3c_s$. This suggests that HQV nucleation requires both spin and mass currents, which is consistent with the spin-mass composite nature of the HQVs. HQVs can be indirectly generated via dissociation of spin vortices that have pure spin circulation~\cite{Seo15,Deveaud12}, but the spin vortices are energetically too costly because of their density-depleted cores.

To quantitatively characterize the energy dissipation by the oscillating magnetic obstacle, we measure the spatial variance of magnetization, $\sigma_M$, at the central region of the condensate. Figure~3 displays the growth of $\sigma_M$ as a function of the stirring frequency $f$. For the weak obstacle, we observe a sudden increase of $\sigma_M$ above a certain critical frequency of $f_c\approx 6$~Hz~\cite{footnote}. This onset behavior indicates the critical generation of spin waves, or magnon excitations, and demonstrates the spin superfluidity of the binary system against external magnetic perturbations. The critical velocity is measured to be $v_c=2\pi A f_c \approx0.7~$mm/s, which is close to the speed of spin sound $c_s$. As $f$ further increases, $\sigma_M$ is observed to be saturated and eventually decrease above $f=16$~Hz. We checked that the stirring time, 1~s, remains in the linear regime with respect to the growth of $\sigma_M$~\cite{Sup}. In Refs.~\cite{Radouani04,Atherton07,Pinsker17}, it was discussed that the excitations are suppressed for a supersonic obstacle due to its finite size.

For the strong obstacle, we observe that $\sigma_M$ starts growing slowly from a low $f>1$~Hz (Fig.~3 inset) and shows a rapid jump at $f_{c,v}\approx 4$~Hz. The preceding growth of $\sigma_M$ indicates the generation of spin waves, while the later rapid increase is due to the HQV shedding, where the magnitude of $\sigma_M$ is significantly enhanced owing to the fully magnetized vortex cores. The two-step growth of $\sigma_M$ reveals that the critical dissipative dynamics evolves from spin-wave emission to HQV shedding in the binary superfluid under the perturbations of the strong magnetic obstacle. The critical velocity for the HQV shedding is measured to be $v_{c,v}\approx0.4~$mm/s, lower than $c_s$. As $f$ increases over 10~Hz, $\sigma_M$ gradually decreases. At the extreme case of $f=50$~Hz, $\sigma_M\approx 0.04$, implying that the generation of HQVs is suppressed. 

\begin{figure} [t]
	\includegraphics[width=8.5cm]{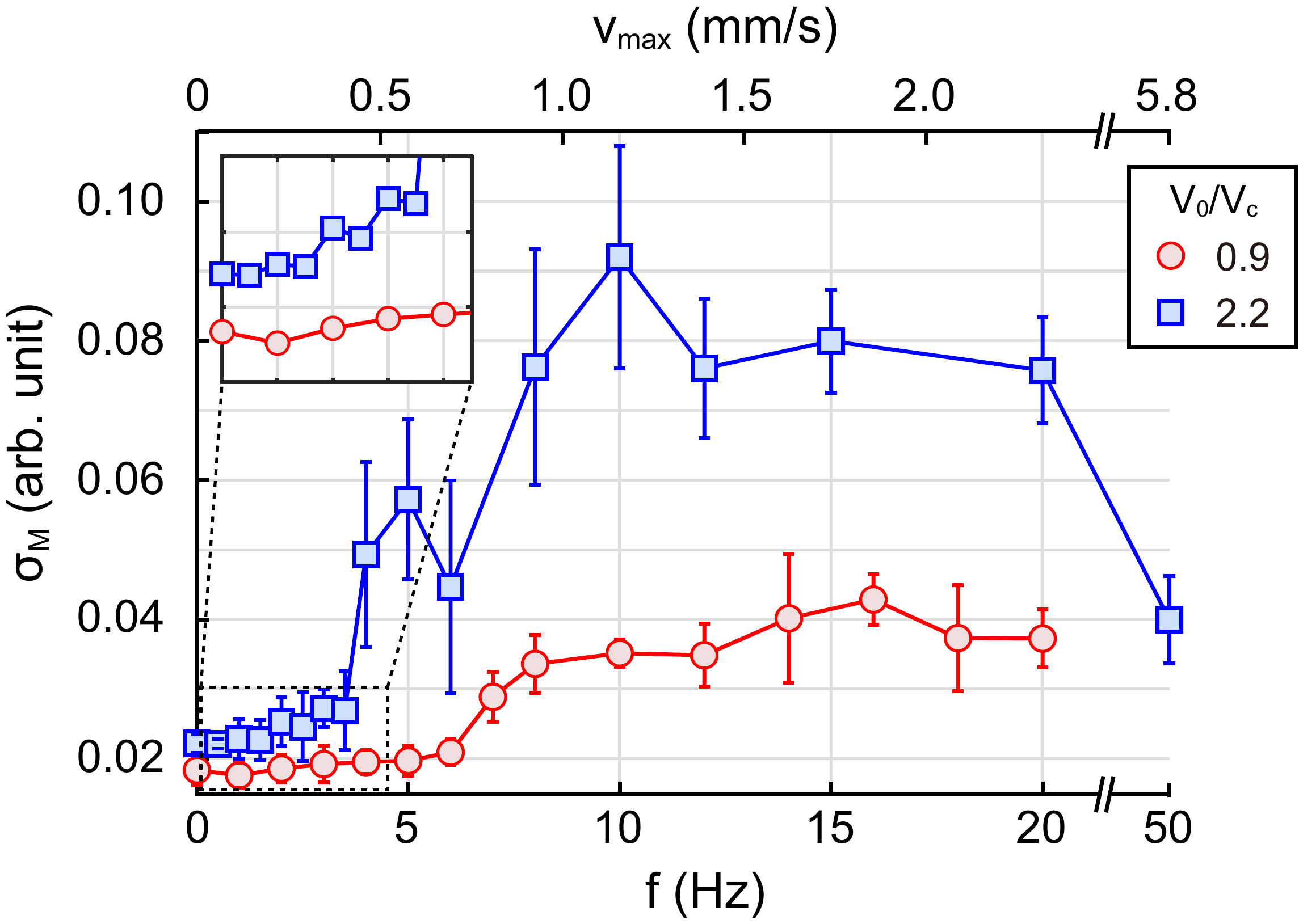}
	\caption{Critical energy dissipation in the binary superfluid. Magnetization variance $\sigma_M$ as a function of the oscillation frequency $f$ for the weak (red circles) and strong (blue squares) obstacles. At the top axis, $v_\text{max}$ denotes the maximum speed of the oscillating obstacle. $\sigma_M$ was measured from the area of $157\times106~\mu$m$^2$ at the central region of the condensate. Each data point is the mean value of five to seven measurements of the same experiment, and its error bar represents their standard deviation. The inset shows an expanded view on the boxed region at low $f$.
	}
\end{figure}

The hierarchy between wave and vortex generations by an oscillating obstacle can be understood from the energy accumulation process for vortex nucleation~\cite{Kwon15-2,Adams00}. When a drag force arises above the critical velocity $v_c$, it gradually accumulates energy in the form of local currents and density compression around the moving obstacle~\cite{Rica92,Adams99}. In the case of oscillating motion, if the amount of the energy accumulated over the oscillation period falls short of the energy cost of a vortex dipole, it is likely to dissipate through wave emission. It was also theoretically shown that the accelerated motion can stimulate the radiation of waves~\cite{Adams00,Mathey16}. Here we note that the relation between phonon emission and vortex shedding was not elucidated in previous stirring experiments with atomic superfluid gases, although the critical velocities were identified by observing a sudden increase of sample temperature~\cite{Ketterle99,Dalibard12,Moritz15}, the onset of a pressure gradient~\cite{Ketterle00}, and the critical vortex shedding~\cite{Anderson10,Kwon15-1,Park18}. In our experiment, spin-wave excitations as well as HQVs are directly detected using the magnetization imaging in the effective 2D sample ($\xi_s>R_z$), which allows one to decipher the two-step evolution of the critical dissipative dynamics.

\begin{figure}[t]
	\includegraphics[width=8.5cm]{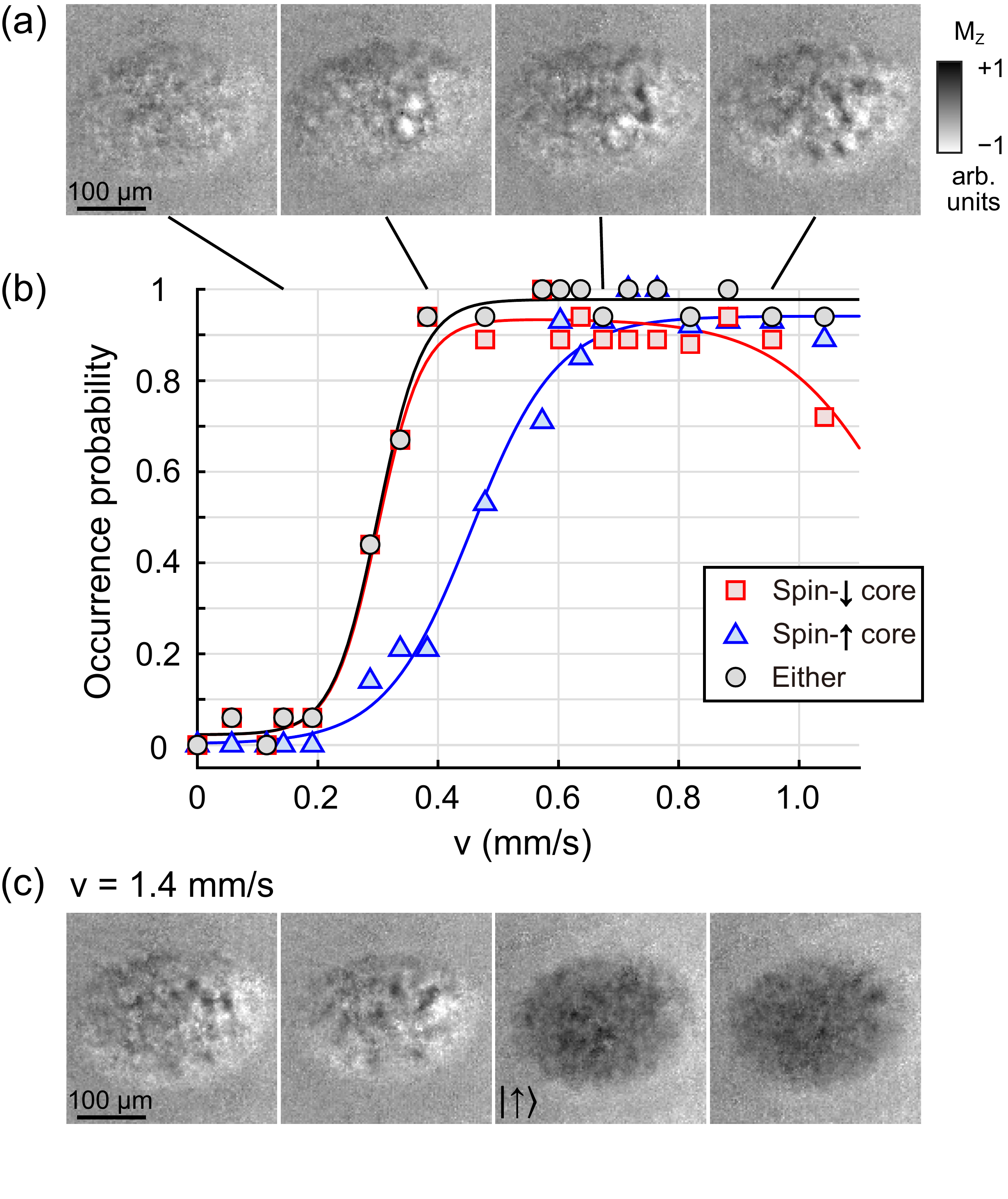}
	\caption{Critical HQV shedding. (a) Magnetization images of the condensate after a linear sweep of the strong obstacle for various obstacles' moving velocity $v$. (b) Occurrence probabilities for spin--$\downarrow$-core HQV (red squares), spin--$\uparrow$-core HQV (blue triangles), and either of them (black circles) as functions of $v$. The probabilities for each $v$ were obtained from 14 to 18 measurements of the same experiment. The solid lines denote a guide for the eyes to each data set based on the sigmoid functions.  (c) Examples of the magnetization images for $v = 1.4$~mm/s. Only spin--$\uparrow$-core HQVs appear. The two images on the right side are the images of the spin--$\uparrow$ component for the same stirring condition. 
	}
\end{figure}

There are two types of HQVs according to the core magnetization, and it is an intriguing query which one is more favorable to be nucleated for the given magnetic obstacle. To examine the detailed aspects of the critical HQV shedding, we perform a modified experiment, where  the strong obstacle with $V_0/V_c \approx 2.2$ is translated at the central region of the condensate by a fixed distance $\approx57~\mu$m with a constant velocity $v$ to shed a few pairs of vortices. In Fig.~4(a), representative magnetization images of the condensate after the linear sweep of the obstacle are shown for various velocities $v$. As $v$ increases, we observe that the HQV shedding dynamics develops in four stages: (i) no excitation arises in the sample, (ii) a HQV dipole with spin--$\downarrow$ core begins to shed above a critical velocity, (iii) a HQV dipole with spin--$\uparrow$ core is also created, and (iv) many HQVs of both types are irregularly generated. The first-shed HQV dipole has cores of the same magnetization as the spin polarization induced by the obstacle. 

The occurrence probability $P_{\uparrow(\downarrow)}$ of the HQVs with spin--$\uparrow(\downarrow)$ core is plotted in Fig.~4(b) as a function of $v$. The onset of vortex generation occurs with the spin--$\downarrow$-core HQVs at $v \approx 0.3~$mm/s, which is slightly smaller than the measured $v_{c,v}$ in Fig.~3, probably due to the difference of the obstacle's motion. We find that HQVs with spin--$\downarrow$ core are always present when spin--$\uparrow$-core HQVs appear at $v \leq 0.4~$mm/s, which implies that the shedding of spin--$\uparrow$-core HQVs requires higher $v$, and that $P_\uparrow$ increases more slowly than $P_\downarrow$.  In the supersonic regime, $v> 0.6$~mm/s, the correlation between the two HQV sheddings is weakened, and interestingly, $P_{\downarrow}$ begins to be suppressed prior to $P_{\uparrow}$.  At high $v>1$~mm/s, it was often observed that only the spin--$\uparrow$-core HQVs appeared in the condensate [Fig.~4(c)].

The nucleation of spin--$\uparrow$-core HQVs is notable because the circulation is formed by the spin component which experiences an attractive potential from the magnetic obstacle. The quantum vortex shedding by an attractive obstacle was not observed in previous experiments~\cite{Dalibard12,Moritz15} and the role of the attractive stirrer is still debatable in numerical studies~\cite{Saito11,Mathey16}. To clarify the issue, we carried out the oscillating obstacle experiment with a scalar condensate containing only the spin--$\downarrow$ component, where $V_0/\mu\approx 1.7$ and the optical obstacle acts as an attractive one. We observed that vortices are generated by the oscillating attractive obstacle above a certain critical frequency~\cite{Sup}. The same experiment was also performed with a condensate of the spin-$\uparrow$ component and it was found that the critical velocity of the attractive obstacle is higher than that of the repulsive one with the same potential magnitude $V_0$. This observation seems to be accounted for by the local Landau criterion at the obstacle position~\cite{Adams00} and provides a qualitative explanation of the measured critical velocities in Fig.~4. Nevertheless, it is important to note that the HQV shedding dynamics cannot be fully described as the sum of the two independent vortex shedding processes. For example, no HQVs were created by the weak obstacle, whereas a penetrable moving obstacle can generate a vortex dipole in a single-component condensate~\cite{Kwon15-2}. Note that HQVs have short-range interactions for different core magnetizations and they are also dynamically coupled to magnons~\cite{Seo16}.

In conclusion, we have studied the critical dissipative dynamics in an antiferromagnetic spinor BEC by moving a magnetic obstacle. The onset of spin-wave excitations was observed for the weak obstacle, directly probing the spin superfluidity of the binary superfluid. The critical HQV shedding was demonstrated with the strong obstacle and the two-step evolution of the critical dissipative dynamics provided insight on the hierarchy between wave emission and vortex generation in the superfluid. An interesting extension of this work is to investigate the spinor superfluid near the quantum critical point with zero quadratic Zeeman energy. Spin superfluidity was predicted to vanish due to the full recovery of spin rotation symmetry~\cite{Kim17} and novel topological objects such as merons and skrymions may exist stably~\cite{Choi12,Blakie20}.

\begin{acknowledgments}
This work was supported by the Samsung Science and Technology Foundation (SSTF-BA1601-06), the National Research Foundation of Korea (NRF-2018R1A2B3003373, NRF-2019M3E4A1080400), and the Institute for Basic Science in Korea (IBS-R009-D1).
\end{acknowledgments}

\clearpage
\newpage

\begin{center}
\textbf{\large Supplemental Material}
\end{center}

\setcounter{equation}{0}
\setcounter{figure}{0}
\setcounter{table}{0}
\makeatletter
\renewcommand{\theequation}{S\arabic{equation}}
\renewcommand{\thefigure}{S\arabic{figure}}
\renewcommand*{\bibnumfmt}[1]{[S#1]}
\renewcommand{\thesubsection}{A\arabic{subsection}}

\subsection*{Magnetic obstacle}

\begin{figure}[b]
	\includegraphics[width=8.5cm]{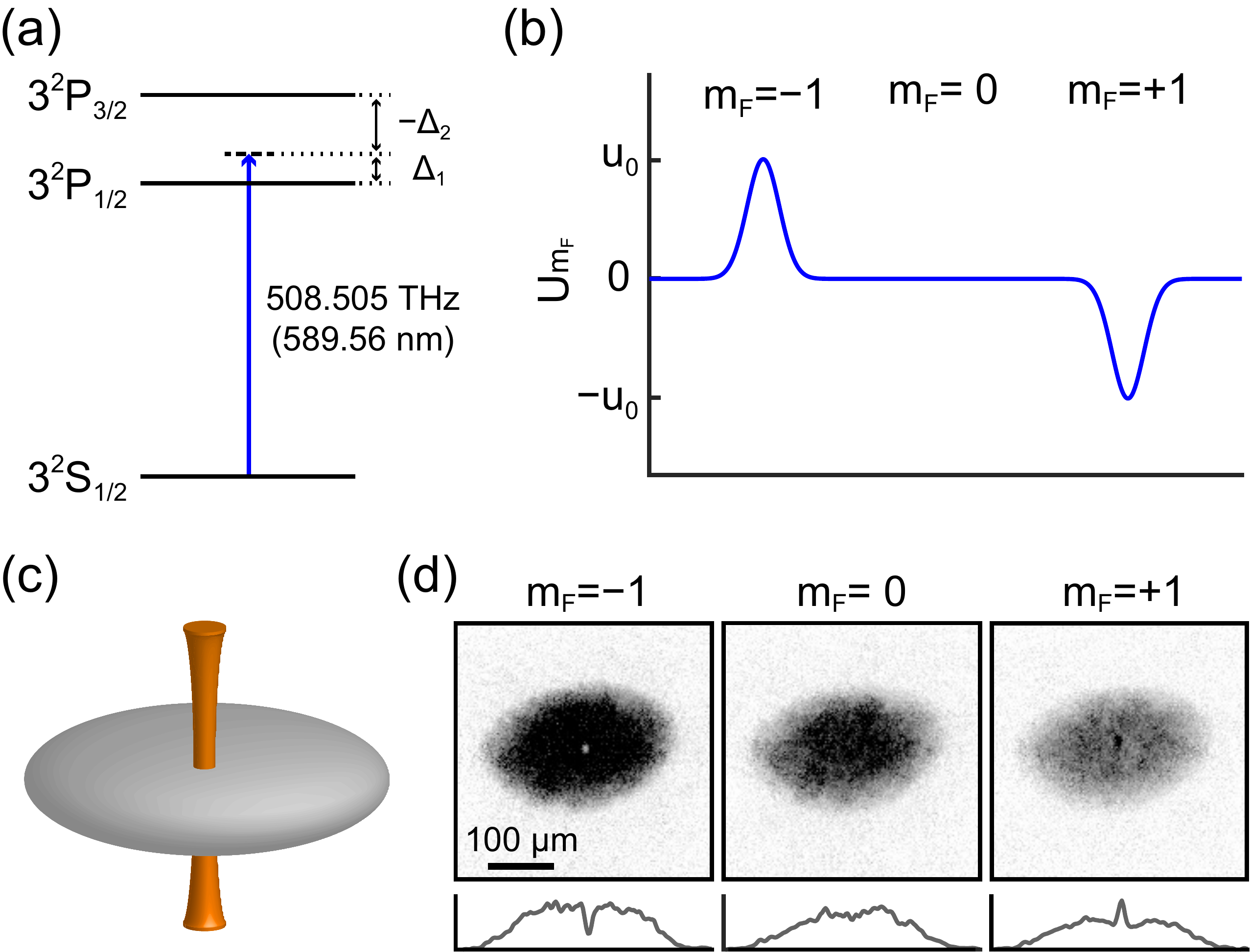}
	\caption{Spin-dependent optical potential for $^{23}$Na. (a) $D$ line doublet of $^{23}$Na with denoting the optical frequency conditions for the magnetic obstacle laser beam. $\Delta_{1,2}$ indicates the frequency detuning from the $D_{1,2}$ transition line. (b) Schematic of the dipole potential $U_{m_F}$ of the laser beam with $\sigma^-$-polarization. The frequency detuning condition of $\Delta_1=-\Delta_2/2= 2\pi\times 172$~GHz yields $U_{-1}=-U_{1}$ and $U_0=0$. $u_0 = (3\pi c^2 \Gamma I)/(8\omega_0^3|\Delta_1|)$.  (c) The focused laser beam penetrates the condensate to form a magnetic obstacle. (d) {\it In-situ} absorption images of the condensate in each spin state. In the bottom, the density profiles along the horizontal line crossing the obstacle position are shown.
}
\end{figure}

For $^{23}$Na in the $F$=$1$ hyperfine ground state, the dipole potential generated by a laser beam is given as 
\begin{equation}
U_{m_F}({\bf r})=\frac{3\pi c^2 \Gamma}{2\omega_0^3} I({\bf r})\Big(\frac{1-g_F m_F P}{3\Delta_{1}}+\frac{2+g_F m_F P}{3\Delta_{2}}\Big),
\end{equation}
where $c$ is the speed of light, $\omega_0$ is the resonance frequency for the $3^2$S$\rightarrow$$3^2$P transition, $\Gamma$ is the decay rate of the excited state, $I({\bf r})$ is the intensity of the laser beam, $g_F=-\frac{1}{2}$ is the Land{\' e} $g$-factor, $m_F=0,\pm1$ is the projection of $F$ on the quantization axis set by the laser beam propagation direction, $P=0, \pm1$ for $\pi$-- and $\sigma^{\pm}$--polarization, and $\Delta_{1,2}$ is the frequency detuning of the laser beam with respect to the $D_{1,2}$ transition line~\cite{Ovchinnikov00}. Here, the hyperfine structures of the excited state are neglected, assumed that their gaps are small enough compared with the frequency detunings $\Delta_{1,2}$.

In the experiment of the main text, we used a 589-nm near-resonant laser beam to produce a magnetic obstacle for $^{23}$Na~\cite{Kim20}. The frequency of the laser beam was set to have $\Delta_1=-\Delta_2/2$, providing $U_{-1}=-U_1$ and $U_0=0$ regardless of $P$ [Fig.~S1(a) and (b)]. Such antisymmetric potentials can be implemented as a magnetic obstacle by focusing and penetrating the laser beam to the condensate [Fig.~S1(c)]. The sign of $U_{\pm1}$ can be inverted by changing the sign of $P$. In our experiment, we used the magnetic obstacle beam that is repulsive for the $m_F$=$1$ state and attractive for the $m_F$=$-1$ state. The potential magnitude $V_0$ was calibrated from the {\it in-situ} density profiles of the spin components [Fig.~S1(d)].

\vspace{0.2in}

\begin{figure}[h]
	\includegraphics[width=7.8cm]{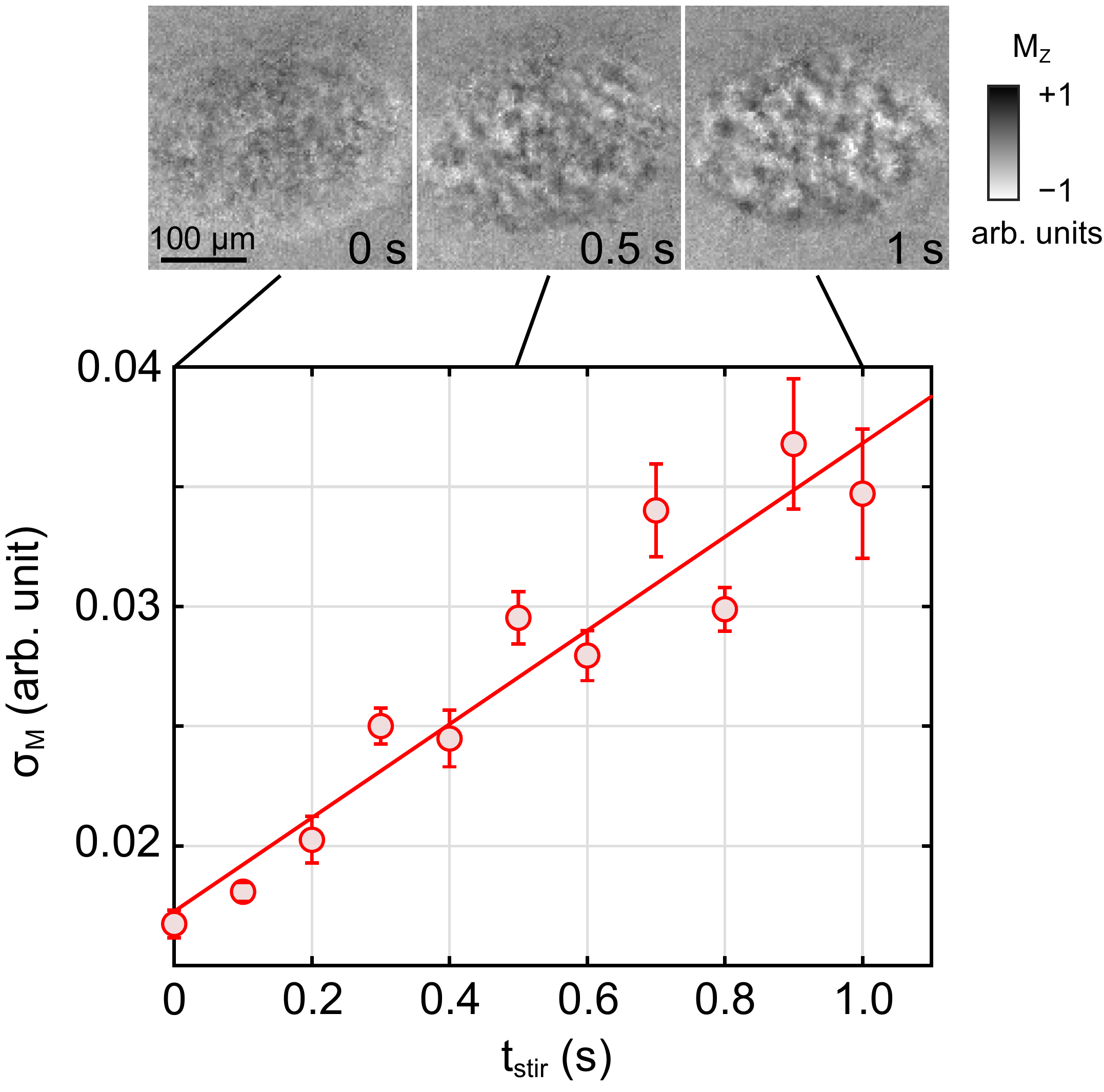}
	\caption{Evolution of $\sigma_M$ as a function of the stirring time $t_\text{stir}$. The weak magnetic obstacle of $V_0/V_c \approx 0.9$ was employed and the stirring frequency was set to be $f=10$~Hz, which is above the critical frequency $\approx 6$~Hz (Fig.~3). $\sigma_M$ was found to increase linearly with $t_{\text{stir}}$ up to 1~s. Each data point was obtained from five measurements of the same experiment and its error bar indicates their standard deviation. In the upper row, the magnetization images of the condensate are displayed for $t_\text{stir}=0$, 0.5, and 1~s, respectively.
}
\end{figure}

\subsection*{Vortex generation by an attractive obstacle}

\begin{figure}[t]
	\includegraphics[width=7.6cm]{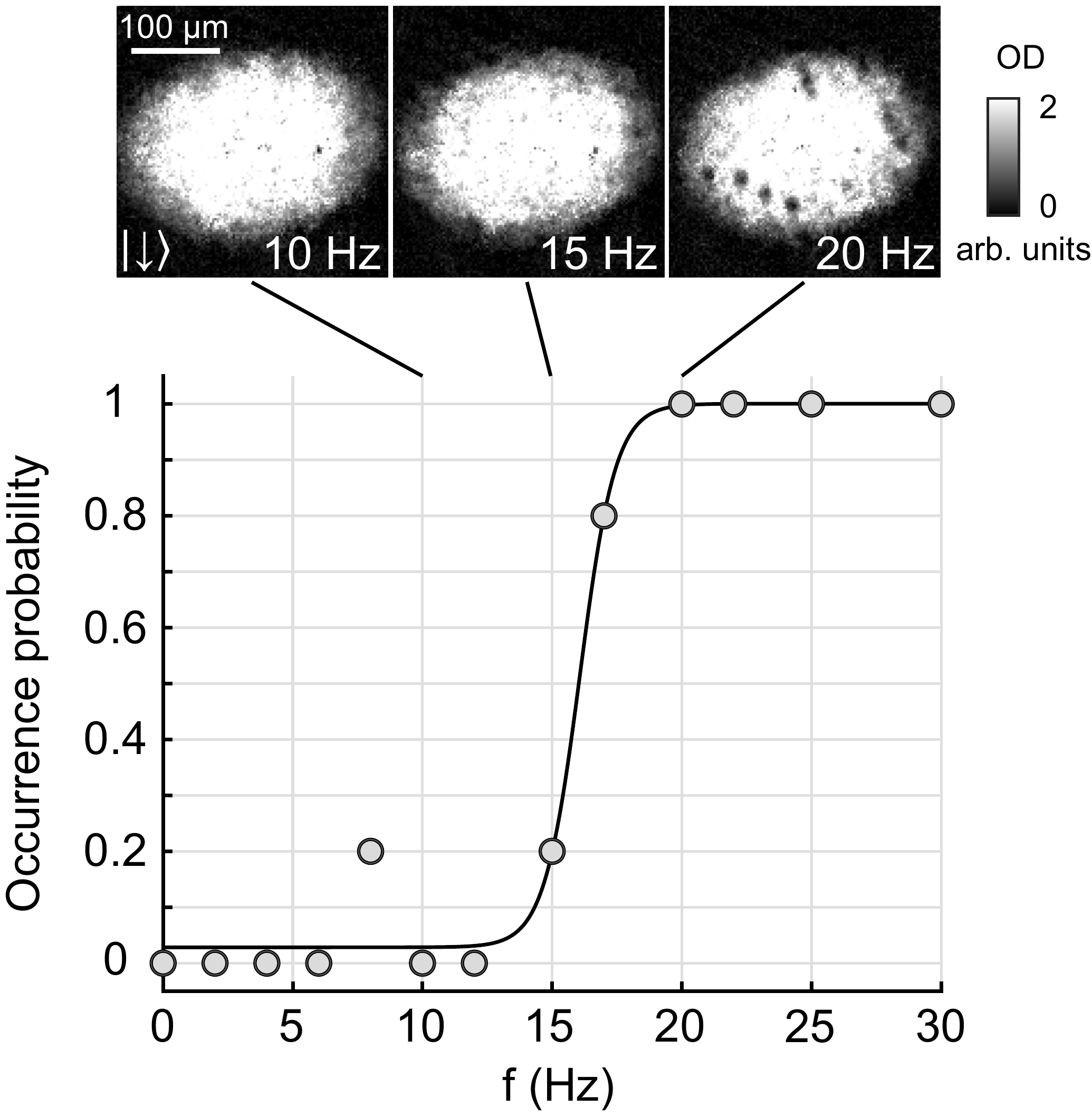}
	\caption{Vortex occurrence probability for an oscillating attractive obstacle as a function of the stirring frequency $f$. The solid line denotes a sigmoidal function fit to the data. Each data point was obtained from five measurements of the same experiment. In the upper row, the optical density images of the condensate are shown for $f=10$, 15, and 20~Hz, respectively.
}
\end{figure}

In order to address the question whether quantum vortices can be generated by a moving attractive obstacle, we performed the same stirring experiment with a condensate prepared to contain only the spin--$\downarrow$ component, where the magnetic obstacle acted as an attractive obstacle. As in the main experiment, the obstacle beam sinusoidally oscillates along a linear path for 1~s with $2A\approx 37~\mu$m. The obstacle strength is $ V_0/\mu_\downarrow \approx 1.7$, where $\mu_\downarrow=gn$ is the chemical potential of the condensate. We indeed observed that quantum vortices were generated in the condensate by the oscillating attractive obstacle above a certain stirring frequency~(Fig.~S3)~\cite{Saito11,Mathey16}. In previous stirring experiments using attractive optical obstacles~\cite{Dalibard12,Moritz15}, vortex generation was not observed and it was attributed to the small size of the obstacles. In our experiment, the $1/e^2$ radius of the obstacle was $\approx 35\xi_n$, where $\xi_n$ is the density healing length of the condensate.  In Fig.~S3, we display the occurrence probability $P(v)$ for quantum vortices in the condensate for various stirring frequencies $f$. From $P(f)=0.5$, the threshold frequency was estimated to be $\approx 16$~Hz, where the maximal speed of the oscillating obstacle is $\approx 1.9$~mm/s, corresponding to about 55\% of the speed of sound in the condensate. For comparison, we carried out the same stirring experiment with a condensate of the spin--$\uparrow$ component, where the same magnetic obstacle acted as a repulsive one, and observed that the threshold frequency of the vortex generation is less than $16$~Hz. This indicates that the critical velocity against the attractive obstacle is higher than that against the repulsive obstacle.


\begin{references}
\bibitem{footnote} $f_c$ is much higher than the spin dipole oscillation frequency of the trapped sample, which is given by $\approx 0.21 \omega_x  = 2\pi \times 1.2$~Hz~\cite{Ferrari18}.
\end{references}
\end{document}